\documentclass[twocolumn,showpacs,preprintnumbers,amsmath,amssymb,prl]{revtex4}
\usepackage{graphicx}% Include figure files
\usepackage{dcolumn}% Align table columns on decimal point
\usepackage{bm}% bold math
\begin{document}
\newcommand{\unity}{\ensuremath{{\rm 1 \negthickspace l}{}}}
%\preprint{APS/123-QED}
\title{Optimal Control of Coupled Josephson Qubits}

\author{A.~Sp{\"o}rl} 
\author{T.~Schulte-Herbr{\"u}ggen}
\email{tosh@ch.tum.de} 
\author{S.J.~Glaser} 
\affiliation{Department of Chemistry, Technical University Munich, 
Lichtenbergstrasse 4, 85747 Garching, Germany.}
\author{V.~Bergholm}% 
\affiliation{Materials Physics Laboratory, POB 2200 (Technical Physics) FIN-02015 HUT, Helsinki University of Technology, Finland.}  
\author{M.J.~Storcz} \author{J.~Ferber}
\author{F.K.~Wilhelm} \email{wilhelm@theorie.physik.uni-muenchen.de}
\affiliation{Physics Department, ASC, and CeNS,
Ludwig-Maximilians-University, Theresienstr.\ 37, 80333 Munich, Germany.\phantom{.}}
%\author{N.~Khaneja}
%\affiliation{Division of Applied Sciences, Harvard University, Cambridge MA02138, USA.}
\date{\today}
\begin{abstract}
\begin{center}
{\em This paper is dedicated to the memory of Martti Salomaa.}\\[1mm]
\end{center}
Quantum optimal control theory is applied to 
two and three coupled Josephson charge qubits. It
is shown that by using shaped pulses a {\sc cnot} gate can be obtained
with a trace fidelity $> 0.99999$ for the two qubits, and even when including
higher charge states, the leakage is below $1\%$.
Yet, the required time is only a fifth of the pioneering experiment
\cite{Nak03} for otherwise identical parameters.  
The controls have palindromic smooth time courses
representable by superpositions of a few harmonics. We outline schemes
to generate these shaped pulses such as simple network synthesis.  The
approach is easy to generalise to larger systems as shown by a fast
realisation of {\sc Toffoli}'s gate in three linearly coupled charge
qubits.  Thus it is to be anticipated that this method will find
wide application in coherent quantum control of systems with 
finite degrees of freedom whose dynamics are Lie-algebraically closed. 

%%%%%%%%%%%%%%%%%%%%

%PACS numbers may be entered using the \verb+\pacs{#1}+ command.
\end{abstract}
%Superconducting Devices, Pulse sequences in NMR
%Quantum Computing, Single electron devices


\maketitle 
In view of
Hamiltonian simulation and quantum computation recent years have seen
an increasing amount of quantum systems that can be coherently
controlled. Next to natural
microscopic quantum systems, a particular attractive candidate for
{\em scalable} setups are superconducting devices based on Josephson
junctions \cite{MSS01}. Due to the ubiquitous bath degrees of freedom in the
solid-state environment, the time over which quantum coherence can be
maintained remains limited, although significant progress has been 
achieved \cite{Bertet04,Astafiev04}. 
Yet, it is a challenge how to produce accurate quantum gates,
and how to minimize their duration such that the number of possible
operations within $T_2$ meets the error correction threshold.
Concomitantly, progress has been made in applying optimal control
techniques %\cite{Pont64} 
to steer quantum systems \cite{Sam90} in a
robust,  relaxation-minimising \cite{KLG} % KG04a} 
or time-optimal way \cite{KBG}.  
Spin systems are a particularly powerful paradigm of
quantum systems \cite{Science98}: under mild conditions they are fully controllable,
{\em i.e.}, local and universal  quantum gates can be implemented.  In $N$
spins-$\tfrac{1}{2}$ it suffices that ({\em i}) all spins can be
addressed selectively by {\em rf}-pulses and ({\em ii}) that the spins
form an arbitrary connected graph of weak coupling interactions.  The
optimal control techniques of spin systems can be extended to
pseudo-spin systems, such as charge or flux states in superconducting
setups, provided their Hamiltonian dynamics can be approximated to
sufficient accuracy by a closed Lie algebra, {\em e.g.}, in a system
of $N$  qubits $\mathfrak{su}(2^N)$.

%\section{Hamiltonian Dynamics of Coupled Charge Qubits}

As a practically relevant and illustrative example, we consider two capacitively coupled charge
qubits controlled by DC pulses as in Ref.\  \cite{Nak03}. The infinite-dimensional Hilbert
space of charge states in the device can be projected to its
low-energy part defined by zero or one excess charge on the respective
islands \cite{MSS01}. Identifying these charges as pseudo-spin states, the
Hamiltonian can be written as 
$H_{\rm tot}=H_{\rm drift}+H_{\rm control}$, where the drift or static 
part reads (for the constants see caption to Fig.~1)
%%%%%%%%%%%
% \begin{equation}
% \begin{split}
%H_{\rm drift} = &-\left(\frac{E_m}{4} + \frac{E_{c1}}{2}\right)
%                (\sigma_z^{(1)}\otimes \unity) -  \frac{E_{J1}}{2}
%                (\sigma_x^{(1)}\otimes \unity) \\\label{eq:drift}
%                &-\left(\frac{E_m}{4} + \frac{E_{c2}}{2}\right)
%                (\unity \otimes \sigma_z^{(2)}) -  \frac{E_{J2}}{2}
%                (\unity \otimes \sigma_x^{(2)})\\ &+ \frac{E_m}{4}
%                (\sigma_z^{(1)}\otimes\sigma_z^{(2)})\quad ,
%\vspace{-9mm}
%\end{split}
%\end{equation}
\begin{eqnarray}
H_{\rm drift} & = & -\left(\frac{E_m}{4} + \frac{E_{c1}}{2}\right)
                (\sigma_z^{(1)}\otimes \unity) -  \frac{E_{J1}}{2}
                (\sigma_x^{(1)}\otimes \unity) \nonumber \\
                & &{} -\left(\frac{E_m}{4} + \frac{E_{c2}}{2}\right)
                (\unity \otimes \sigma_z^{(2)}) -  \frac{E_{J2}}{2}
                (\unity \otimes \sigma_x^{(2)}) \nonumber \\ 
                & &{} + \frac{E_m}{4} (\sigma_z^{(1)}\otimes\sigma_z^{(2)})\quad , \label{eq:drift}
\end{eqnarray}
%%%%%%%%%%%
while the controls can be cast into
%%%%%%%%%%%
\begin{equation}
\begin{split}
H_{\rm control} = &\left(\frac{E_m}{2}n_{g2} + {E_{c1}}n_{g1}\right)
                        (\sigma_z^{(1)}\otimes \unity)\\ +
                        &\left(\frac{E_m}{2}n_{g1} +
                        {E_{c2}}n_{g2}\right)  (\unity \otimes
                        \sigma_z^{(2)})\quad . \label{eq:control}
\end{split}
\end{equation}
%%%%%%%%%%%
The control amplitudes $n_{g\nu}$, $\nu=1,2$ are gate charges %%% NOTE: in units of Cooper pairs
controlled by external voltages {\em via}
$n_{g\nu}=V_{g\nu}C_{g\nu}/2e$. They are taken to
be piece-wise constant in each time interval $t_k$. This pseudo-spin 
Hamiltonian motivated by Ref.\ \cite{Nak03} also applies to other
systems such as double quantum dots \cite{Hayashi03} and 
Josephson flux qubits \cite{Majer05}, although in the latter case the
controls are typically {\em rf}-pulses.

In a time interval $t_k$ the system thus evolves
under $H_{\rm tot}^{(k)} = H_{\rm drift} + H_{\rm control}^{(k)}$.
The task is to find a sequence of control amplitudes for the
times $t_1,t_2,\dots, t_k, \dots, t_N$ such as to maximise a
quality function, here the overlap with
the desired quantum gate or element of an algorithm $U_{\rm target}$. 
Moreover, for the decomposition of
$U_{\rm T} = e^{-it_N H_N} e^{-it_{N-1}H_{N-1}} \cdots e^{-it_k H_k}
\cdots e^{-it_1 H_1}$
into
available controls $\{H_\nu\}$ to be timeoptimal,
$T:=\sum_{k=1}^N t_k$ has to be minimal. 
The gate fidelity is unity, if  
${||U_T - U_{\rm target}||}_2 = 0 =
{||U_T||}_2^2 + {||U_{\rm target}||}_2^2  - 2 {\rm Re\; tr} 
\{U_{\rm target}^\dagger U_T\}$.  Maximising 
${\rm Re\; tr} \{U_{\rm target}^\dagger U_T\}$ for fixed $T$ can readily be solved by optimal
control: Let $h\big(U(t_k)\big) :=
{\rm Re}\;{\rm tr}\{\lambda^\dagger(t_k) (-i (H_{\rm d}+\sum u_\nu
H_\nu))U(t_k)\}$ with the Lagrange-type adjoint system $\lambda(t)$ following the
equation of motion $\dot{\lambda}(t) = -i (H_{\rm d} + \sum u_\nu H_\nu)
\lambda(t)$.  Pontryagin's maximum principle requires
%%%%%%%%%%%%%%%%%%%
\begin{figure}[Ht!]
\includegraphics[width=3.25in]{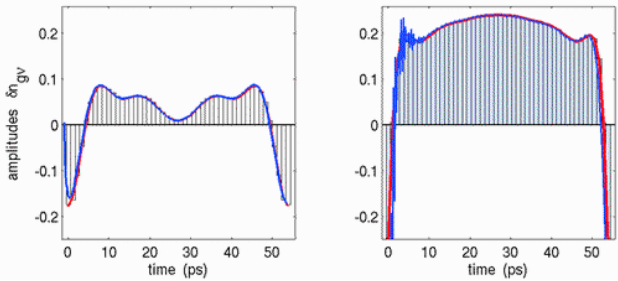}
\caption{(Color online) Fastest charge level controls obtained for realising a {\sc
cnot}-gate on  a pair of coupled charge qubits 
(left part: control qubit, right part: working qubit). The total gate charges
for the qubits are $n_{g\nu}=n_{g\nu}^{(0)}+\delta n_{g\nu}$ with $\nu=1,2$.
Here, $n_{g1}^{(0)}=0.24$, $n_{g2}^{(0)}=0.26$ and the qubit energies
$E_{c1}/h=140.2$ GHz, $E_{c2}/h=162.2$ GHz, $E_{J1}/h=10.9$ GHz,
$E_{J2}/h=9.9$ GHz, and $E_m/h=23.0$ GHz were chosen in accordance with
the experimental values given in \cite{Nak03}.  The 50 piecewise
constant controls are shown as bars; 
the trace fidelity is 
$\tfrac{1}{N}\big|{\rm tr} \{U_{\rm target}^\dagger U_T\}\big| > 1-10^{-9}$.
Red lines give the analytic curves in Eqn.~\ref{envelope}, while the blue ones superimposed
show a pulse actually synthesised by an LCR-filter (see a later section and Fig.~3).
}
\end{figure}
%%%%%%%%%%%%%%%%%%%
${\partial h}/{\partial u_\nu} \equiv {\rm Re}\;{\rm
tr}\{\lambda^\dagger (-i H_\nu)U\} = 0$  
thus allowing to implement a gradient-flow based recursion.
%%building upon the principles of \cite{Bro88} extended to unitary control in \cite{Science98}.  
For the amplitude of the $\nu^{\rm th}$ control in
iteration $r+1$ at time interval $t_k$ one finds with $\varepsilon$ as
a suitably chosen step size \; $n^{(r+1)}_{g\nu}(t_k) =
n^{(r)}_{g\nu}(t_k) + \varepsilon \tfrac{\partial
h^{(r)}(t_k)}{\partial n^{(r)}_{g\nu}(t_k)}$ as derived in
Refs.~\cite{KG04b,PRApreprint}. Here $T$ is the shortest
time allowing for a given fidelity numerically. 

We now turn to the discussion of our numerical results. We have used
parameter values from the experiment \cite{Nak03}. Variation of  these
values should change details of the result, but not its overall
structure.  Fig.~1 shows the fastest decompositions obtained by numerical optimal control
for the {\sc cnot} gate into
evolutions under available controls (Eqns.~1 and 2). In contrast to
the $250$ ps in Ref.~\cite{Nak03}, $T=55$ ps suffice to get
${||U_T-U_{\rm target}||}_2 = 5.3464 \times 10^{-5}$ corresponding to
a trace fidelity of 
$\tfrac{1}{N}\big|{\rm tr} \{U_{\rm target}^\dagger U_T\}\big| > 1-10^{-9}$.

Beyond the efficient and accurate
implementation, this result provides physical insight:
our pulse essentially accomodates all terms of the standard
{\sc cnot} pulse sequence for this coupling \cite{MSS01} such that
different terms in the total Hamiltonian act in parallel instead
of sequentially. 
For a {\sc cnot}, the duration $T=55$ ps has to accomodate at least a
$\tfrac{\pi}{2}$ rotation under the coupling Hamiltonian ($\tfrac{1}{2}\sigma_z\otimes\sigma_z$)
lasting 21.7 ps concomitant to two
$\tfrac{\pi}{2}$ $x$-rotations under the second of the drift components
($\tfrac{1}{2}\sigma_x^{(\nu)}$ with $\nu=1,2$)
requiring 22.9 and 25.3 ps, respectively. Thus, unlike 
in NMR, the time scales of local and non-local interactions are comparable.
Assume in a limiting simplification
that {\em two} $\tfrac{\pi}{2}$ $x$-pulses are required, the total length cannot
be shorter than $50.6$ ps. Our solution is close to this infimum.
Note that a duration of $T=55$ ps also implies that the trajectory
of the coherent evolution does not have to be a geodesic in the Weyl chamber
(compare ref.~\cite{KBG}), as shown in the supplement Fig.~5. Moreover,
the evolution times for single components do not add up, indicating
parallel evolution of different interactions.

The supplementary material illustrates how the sequence of
controls (Fig.~1) acts in a quasi-continuous way on specific input
states:  Suppl.~Fig.~1 gives the evolution of a product state,  $
|\Theta\rangle = |0\rangle|0\rangle$. The representation of the
reduced states in their local Bloch spheres shows how the control
qubit undergoes a closed loop, while the working qubit is inverted as
expected.  As demonstrated in Suppl.~Fig.~2, a maximally entangled
Bell state $|\psi_+\rangle = \tfrac{1}{\sqrt{2}} (|0\rangle|0\rangle +
|1\rangle|1\rangle$  evolves from the centre of the Bloch spheres
(indicating maximal entanglement) into a product state on the
surface of the sphere.
%%%%%%%%%%%%%%%%%%%
\begin{figure*}[Ht!]
\includegraphics[width=5.5in]{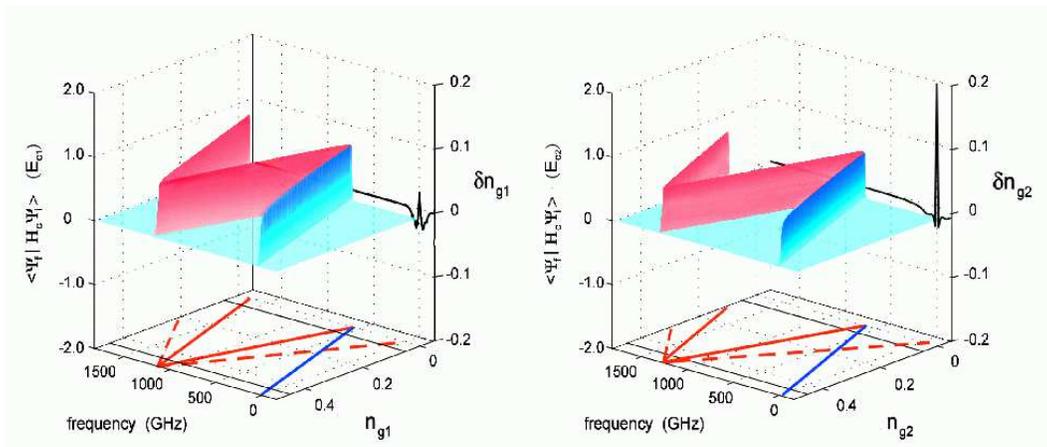}
\caption{(Color online) Spectroscopic explanation of the high quality of the control
sequences of Fig.~1: the spectral overlap of the Fourier-transforms (right walls) of
the controls of Fig.~1 with the energy differences
corresponding to the one-charge transitions into leakage levels
(solid lines on the surface) is small at charges around
$n_{g\nu}=0.2\;\text{with}\;\nu=1,2$. Intensities at allowed (solid lines) 
{\em vs} forbidden transitions (broken lines)
into leakage levels are given in terms of the transition-matrix elements
%with the control Hamiltonian $H_c$ of Eqn.~2 
$\langle \Psi_f | H_c \Psi_i\rangle$
normalized by the charging energies $E_{c1}$
($E_{c2}$) in the 3D representation: 
the working transitions (blue) are of the same probability as the
allowed transitions into leakage levels (red) that have no overlap with the excitation bandwidth
of the pulses, while the forbidden ones are too weak to show up at all.
(details in the text). 
}
\end{figure*}
%%%%%%%%%%%%%%%%%%%

Note that the time course of controls in charge qubits 
turns out palindromic (Fig.~1).  Self-inverse gates ($U^2_{\rm gate} = \unity$)
relate to the more general time-and-phase-reversal symmetry
(TPR) observed in the control of  spin systems \cite{GGSE87}: for
example, any sequence  $e^{-it_x\sigma_x} e^{-it_y\sigma_y}
e^{-it_z\sigma_z}$ is inverted by transposition concomitant to time
reversal $t_\nu\mapsto -t_\nu$ and $\sigma_y\mapsto -\sigma_y$. Since
the Hamiltonians in Eqns.~1-2 are real and symmetric, they will give
the same propagator, no matter whether read forward or backward.

The pulse is not very complicated.  Interestingly, the time-course of
the controls on either qubit ($\nu = 1,2$) can be written as a sum of
$6 (7)$ harmonic functions
\begin{equation}\label{envelope}
n_{g\nu} (t) = \sum\limits_{j=0}^{5(6)} a_\nu(j)
        \cos\big(2\pi \omega_\nu(j)\frac{t}{T}  +
        \phi_\nu(j)\big)\;.
\end{equation}
The constants from Tab.~1 in the supplementary material
give a high accuracy ($\chi^2 = 0.008231;
0.003668$ for  the channels 1 and 2, respectively).

This representation reflects the simplicity and the modest bandwidth
of the pulses obtained. The low bandwidth allows to maintain a high
fidelity even if leakage levels formed from higher charge states of
the qubit system  are taken into account:  
we now explicitly apply the full pulse to the
extended Hamiltonian obtained by mapping the full Hamiltonian
\cite{Nak03} to the subspaces of $-1,\dots,2$ extra charges per island.
The two-qubit
{\sc cnot} gate is then embedded into the group $SU(16)$. Even then
the propagator generated by
the above controls projects well onto the {\sc cnot} gate
giving a trace fidelity $> 0.99$. 
The good result may be astounding at first sight, however, it can be
understood by relating the limited bandwidth to
the matrix elements, which both control the transition rate: while the {\em one-charge}
transitions to the leakage levels like  $|-1\rangle \leftrightarrow
|0\rangle$ and $|2\rangle \leftrightarrow |1\rangle$  are 
allowed by the Josephson coupling,  the {\em two-charge}
transitions like  $|-1\rangle \leftrightarrow |1\rangle$ and $|2\rangle
\leftrightarrow |0\rangle$  are forbidden in terms of the transition probabilities 
$\langle \Psi_f | H_c \Psi_i\rangle$
as can be
seen from Fig.~2. Moreover, note that the charge levels of Fig.~1 are
mostly around $n_g=0.2$ thus contributing to the working transition
$|0\rangle \leftrightarrow |1\rangle$, while the \/`spectral
overlap\/' of the Fourier-transform of the time course of the controls
with energy differences corresponding to the potentially deleterious
one-charge transitions in Fig.~2 is small. Hence there are simple
spectroscopic arguments for the high fidelity obtained by
the controls of Fig.~1.

Furthermore, even the time courses starting out with any of the four
canonical two-qubit basis vectors hardly ever leave the state space 
of the working qubits: at any time the projections onto the leakage space
do not exceed $0.6\,\%$. Choosing initial states from the Bell basis
entails even less leakage.
Note that explicitly taking into account the leakage levels during optimisation
is expected to improve the quality even further.
Thus,  also pseudospin systems like ours,
which involve a low-energy  projection disregarding
leakage levels, can be controlled with high
accuracy and quantum computing is {\em not} strictly limited to 
coupled two-level systems.

Generating these pulse shapes experimentally is a challenging but
possible  task. Note that the length of the pulse is given by the
coupling strength as discussed above and hence can be extended by
lowering the coupling.

In the pertinent time scale, there are no devices comercially available
for generating arbitrary wave forms with the same capabilities as NMR-spectrometers.
High-end commercial pulse generators as well as custom-built ones are
close to the necessary specifications \cite{Kim05,Qin02}. Pulses can be formed by superimposing
short pulses of shapes easy to generate with different heights,
widths, and delays. The two main candidates for this approach are ({\em i})
Gaussian pulses \cite{Hayashi03}, which can be generated at  room
temperature and which run nearly undistorted through the necessary
cryogenic filtering and ({\em ii}) SFQ pulses, which can be generated on
chip (hence avoiding the filters) using ultrafast classical Josephson
electronics \cite{Brock,Crankshaw02}.

We would like to exemplify a well established technique, 
shaping in Laplace space, to generate these pulses. The
idea resembles the approach of femtosecond quantum chemistry: we start
with an input current pulse $I_{\rm in}(t)$ {\em shorter} than the
desired one of a shape which is arbitrary as long as it contains
enough spectral weight at the harmonics necessary for the desired
pulse. Such pulses are readily generated optically or  electrically
and have, without shaping, already been applied under cryogenic
conditions  \cite{Qin02}. This pulse is sent through a discrete
electrical  four-pole, whose transfer function $Z_{\rm 12}$ is
designed such that the desired pulse is found at the output. We have
carried out this idea for a rectangular pulse of length $\tau_r=1.1 {\rm ps}$  
as an input and our two gate pulses as
output. We have  developed a transfer function in Laplace space
$Z_{12} (s)$ by fitting $V_{\rm g} (s)=Z_{12}(s) I_{\rm in}
(s)$ (see Fig.~3). Owing to causality, the poles of $Z_{\rm 12}$ are  either on the
negative real axis or in conjugate pairs of poles on the left half
plane. Each conjugate pair corresponds to an LCR-filter stage whereas
each real pole corresponds to an RC lowpass-filter. It turns out, that
good agreement can be achieved with 8 LCR filters and two
low-pass filters, following the standard rules of circuit synthesis \cite{Rupprecht}.  

The pulses are very close to the desired ones, see Fig.~1, and a trace
fidelity of  94 \% can be achieved for the entire {\sc cnot}.
Clearly, the quality can be further improved with more refined technology.
%%%%%%%%%%%%%%%%%%%%%%%
\begin{figure}[Ht!]
\includegraphics[width=0.7\columnwidth]{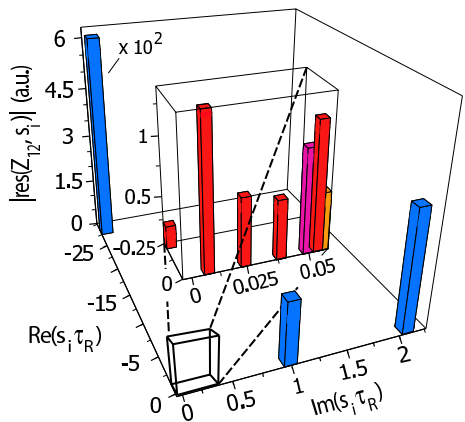}
\caption{(Color online) Characterization of a filter shaping the pulse on the second gate. 
The bars show the poles $s_i$ of the transfer function in the Laplace
plane. Poles outside the negative imaginary axis also lead to the complex conjugate
pole and can be implemented by an LCR-Filter. The height of the bars show the 
modulus of the residue in this pole.  
\label{fig:filtering}}
\end{figure}
%%%%%%%%%%%%%%%%%%%%%%%

The filter as well as the pulse design are ready to accomodate the
experimental necessities. On the one hand, due to unavoidable
fabrication uncertainties, the optimum pulse will look slightly 
different for
each individual pair of qubits.  Realistically, the matrix elements of
the total Hamiltonian Eqs.\, (\ref{eq:drift}), (\ref{eq:control}) first
have to be determined spectroscopically, then our algorithm has to be
run to find the optimum pulse shape. This is done on a regular PC in
ca.\ 30 seconds. Secondly, for adjusting the filtering circuit which
can
be put at room temperature, one has to
take into account the transfer function through the filters of
the cryostat and to the sample. 
This contribution $Z_{\rm sample}$ to the total
transfer function will most realistically be measured using a capacitor of
the same geometrical dimensions of the qubits as a probe. As long 
as this does not block the relevant frequencies, {\em i.e.},\ if the setup
has sufficient bandwidth, $Z_{\rm sample}$ can be accounted for
when adjusting an additional pulse shaping filter 
such that the {\em total} transfer function shapes
the correct pulse. Note, that our method also applies to control by microwave Rabi-type
pulses, where pulse shaping appears to be easier as time scales are usually longer.

Likewise, in a system of three linearly coupled charge qubits, one may
decompose the {\sc Toffoli} gate into experimentally available controls.

%%%%%%%%%%%%%%%%%%%
\begin{figure}[Ht!]
\includegraphics[scale=0.4]{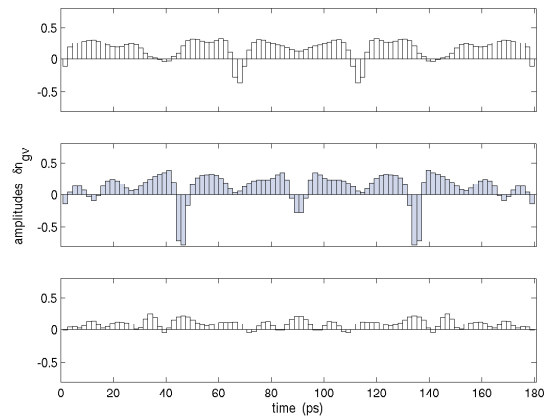}
\caption{Fastest charge level controls obtained for realising a Toffoli-gate on 
a linear chain of charge qubits coupled by nearest-neighbour interactions.
The piecewise constant controls are shown as bars.
The trace fidelity is
$\tfrac{1}{N}\big|{\rm tr} \{U_{\rm target}^\dagger U_T\}\big| > 1-10^{-5}$.
%${||U_T-U_{\rm target}||}_2 = 0.0064149$.
Here, the parameters are $E_{c1}/h=140.2$ GHz, $E_{c2}/h=120.9$ GHz, $E_{c3}/h=184.3$ GHz, $E_{J1}/h=10.9$ GHz,
$E_{J2}=/h9.9$ GHz, $E_{J3}/h=9.4$ GHz,$E_{m1,m2}/h=23$ GHz, $n_{g,1}^0=0.24$, $n_{g,2}^0=0.26$, $n_{g,3}=0.28$.}
\end{figure}
%%%%%%%%%%%%%%%%%%%

This result highlights that due to the comparatively strong qubit-qubit interactions 
in multiqubit setups, the direct generation of three-qubit gates is
much quicker than its compostion into elementary universal gates , {\em e.g.}
decomposing a {\sc Toffoli} into 9 {\sc cnots} in a linear spin chain: 
the speed-up is by a factor of 2.8 compared to 9 of our {\sc cnots} and
by a factor of 13 compared to Nakamura's {\sc cnots} \cite{Nak03}.

This also holds when developing
simple algorithms \cite{Vartiainen03} on superconducting qubit setups:
a minimization
algorithm for searching control amplitudes in coupled Cooper pair boxes
has been
applied in \cite{Niskanen03}, however, in that approach, the  numerical
optimization was restricted to only a few values. 
In Ref. \cite{Rigetti04}, a pulse sequence
generating a {\sc cnot} with fixed couplings has been invented, which uses hard
RF pulses instead of our shaped pulse and turns out to be much longer, thus
leads to serious conflicts with decoherence.

In conclusion, we have constructed pulses for the realization of fast
high-fidelity quantum logic gates in superconducting charge qubits. 
The optimum pulses are always palindromic, owing to the time-reversal
invariance of these pseudo-spin Hamiltonians. 
The simplicity of the pulse shape results in low bandwidth and thus
low leakage to higher states and the setup necessary to generate such 
pulses is of modest complexity. 

%\begin{acknowledgments}
We are indebted to Navin Khaneja for continuous stimulating scientific exchange.
We thank M.\ Mariantoni for extensive discussions on experimental issues,
specifically the transfer functions, as well as 
Y.\ Nakamura, J.M.\ Martinis, A.\ Ustinov, and D.\ van der Weide.
This work was supported in part by {\em Deutsche Forschungsgemeinschaft},
DFG, {\em Schwerpunkt Quanten-Informationsverarbeitung} (SPP 1078: Gl 203/4-2) and SFB 631.
MJS, JF, and FKW acknowledge support of ARDA and NSA through ARO contract 
P-43385-PH-QC. 

%\end{acknowledgments}

\bibliography{draft}

%%%%%%%%%%%%%%%%%%%%%%%%

%%%%%%%%%%%%%%%%%%%%%%%%

%%%%%%%%%%%%%%%%%%%%%%%%

\end{document}

\newcommand{\unity}{\ensuremath{{\rm 1 \negthickspace l}{}}}
%\newcommand{\unity}{\ensuremath{{\rm 1\mkern-4.6mu l}{}}\xspace}

%\preprint{APS/123-QED}

\title{Supplementary Material:\\ Optimal Control of Coupled Josephson Qubits}

\author{A.~Sp{\"o}rl, T.~Schulte-Herbr{\"u}ggen, S.J.~Glaser, V.~Bergholm, M.~Storcz, J.~Ferber, and
F.K.~Wilhelm}
\date{\today}
\maketitle
%%%%%%%%%%%%%%%%%%%
%%%%%%%%%%%%%%%%%%%

\section{Detailed presentation of the pulse and the qubit dynamics}

\begin{figure}[b!]
\includegraphics[scale=.62]{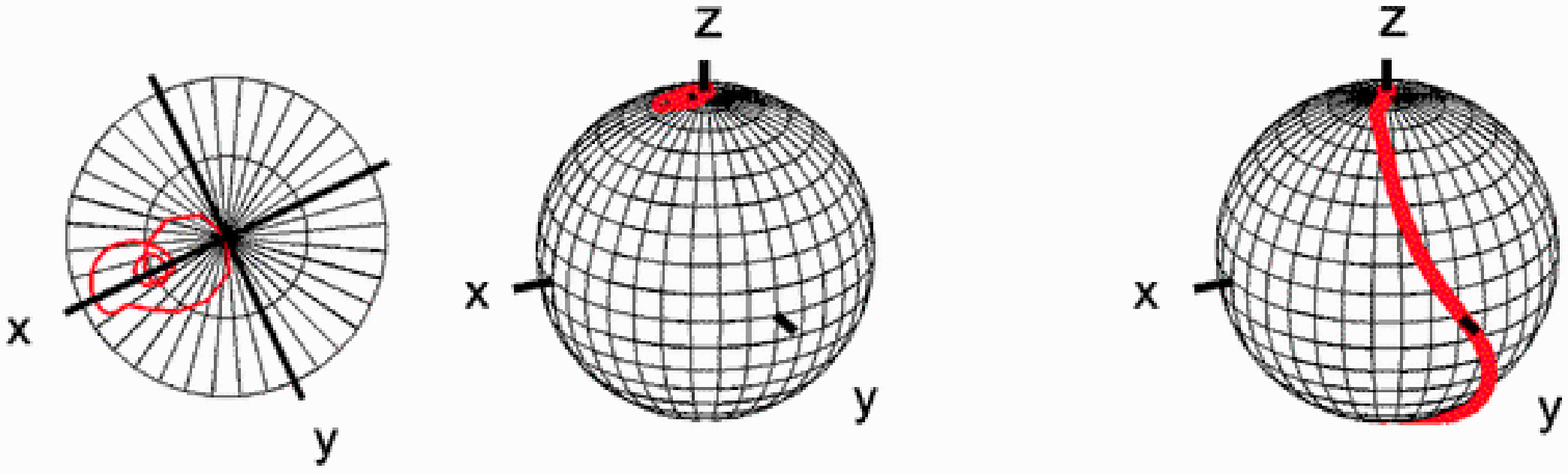}
\caption{ Evolution of the product state $|\Theta(0)\rangle=|0\rangle|0\rangle$ under the optimised 
controls resulting in 
$|\Theta(T)\rangle=|0\rangle|1\rangle$. The evolution $0<t<T$ with $T=55$ ps is represented by the reduced states 
$tr_B|\Theta(t)\rangle\langle\Theta(t)|$ (left sphere) and $tr_A|\Theta(t)\rangle\langle\Theta(t)|$ (right sphere) 
on the respective local Bloch spheres. The blow-up on the left shows the top of the left Bloch sphere.
\label{fig:bloch1}}
\end{figure}

\begin{figure*}[b!]
\includegraphics[scale=.62]{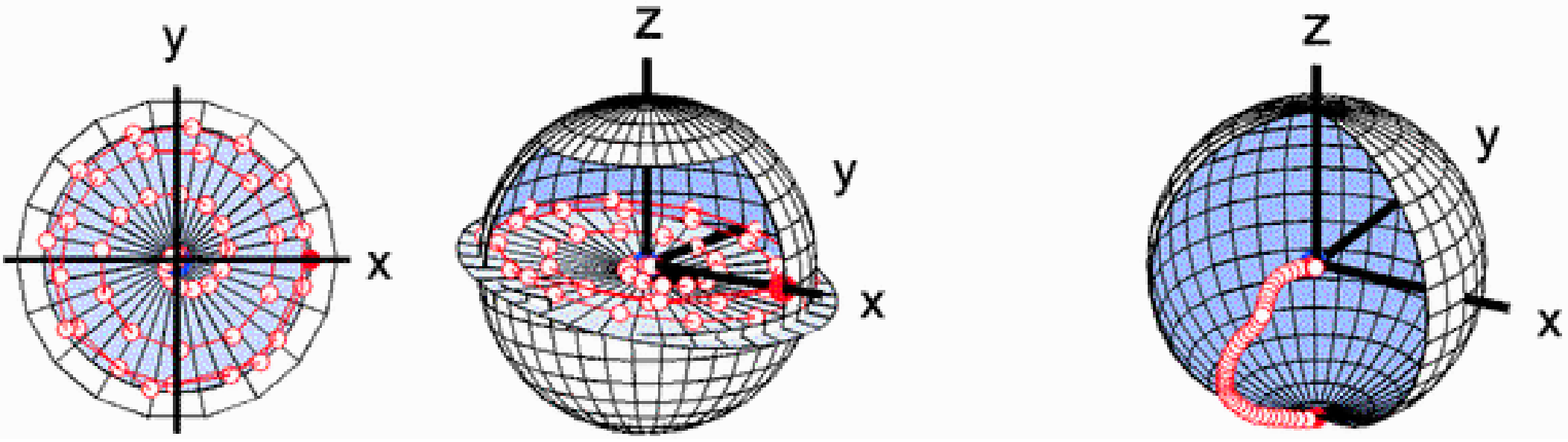}
\caption{Evolution of the Bell state $|\Phi_+\rangle = \tfrac{1}{\sqrt{2}}\left(|00\rangle+|11\rangle\right)$ 
into the final state $\tfrac{1}{\sqrt{2}}\left(|01\rangle+|11\rangle\right)$ (filled red dots). 
The Bell state is maximally entangled and hence has local representations in the centre of the 
respective Bloch spheres, while the final state is a product state represented by points (filled red dots) 
on their surfaces. The projection on the left is a view from the top onto the plane inserted 
into the left Bloch sphere. \label{fig:bloch2}
}
\end{figure*}

\begin{figure*}[b!]
\includegraphics[scale=.62]{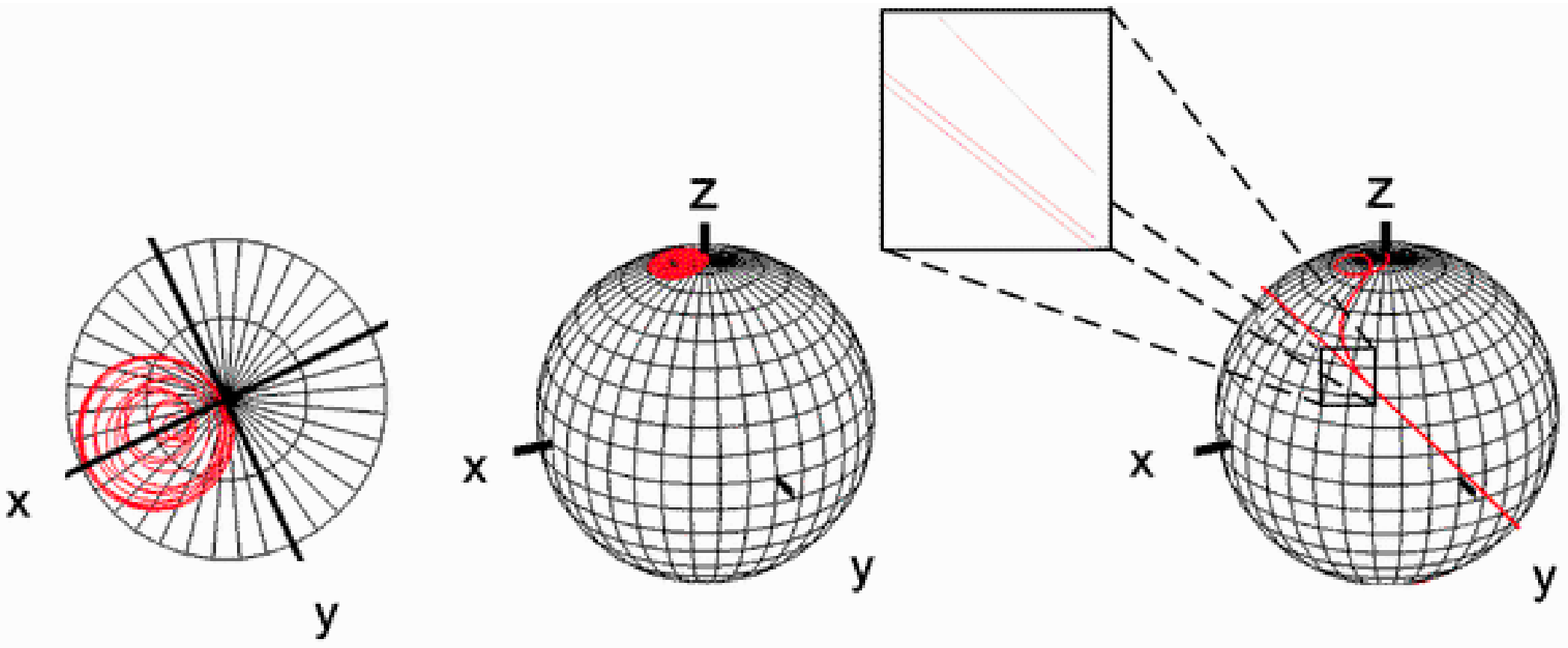}
\caption{For comparison:
same evolution of the product state $|\Theta(0)\rangle=|0\rangle|0\rangle$ as in Fig.~1, but
using the pulse of the experiment Ref. \cite{Nak03}. The evolution $0<t<T$ with $T=255$ ps is represented by the reduced states 
$tr_B|\Theta(t)\rangle\langle\Theta(t)|$ (left) and $tr_A|\Theta(t)\rangle\langle\Theta(t)|$ 
(right) on the respective local Bloch spheres.
The trajectory completes two full circles (see inset)
before reaching its final state near the south pole.
\label{fig:bloch3}}
\end{figure*}

In this supplement, we provide more insights into the numerically obtained pulse and the resulting qubit
dynamics as an additional visualization. Table I contains the precise numerical settings for the 
Fourier decomposition of the pulse shown in Fig. 1 and Eq.\  3 of the paper. 
%%%%%%%%%%%%%%%%%%%
Figure \ref{fig:bloch1} shows the qubit dynamics on the reduced Bloch-spheres of the two qubits,
i.e. the left Bloch spere, belonging to the control qubit, qubit 1, shows the spin projections of the reduced
density matrix, $\rho_1={\rm Tr}_2 \rho$, and vice versa. Even if the two-qubit state is pure, the reduced Bloch vector may be inside the sphere, which hints at entanglement between the qubits, as seen in 
Fig.\ \ref{fig:bloch2}. The experimental pulse shown in Fig.\ \ref{fig:bloch3} requires a much longer trajectory on the sphere. 
%%%%%%%%%%%%%%%%%%%
%%%%%%%%%%%%%%%%%%%
%%%%%%%%%%%%%%%%%%%
%%%%%%%%%%%%%%%%%%%
\begin{table}[h!]
\caption{Parameters giving the envelope to the
control amplitudes $n_{g\nu}(t)$ for the two qubits $\nu=1,2$
as in Eqn.~3 in the main text.
$T=55$ ps. The goodness of fit is determined by
$\chi^2 = 0.008231$ for qubit 1 and $\chi^2 = 0.003668$ for qubit 2. }
\begin{ruledtabular}
%\begin{tabular}{.......} %using package dcolumn
\begin{tabular}{ccccccc}
$j$&$a_1(j)$& $\omega_1(j)$& $\phi_1(j)$& $a_2(j)$ &$\omega_2(j)$ & $\phi_2(j)$\\[1mm]
\hline\\[-2mm]
0  &$-$4.4647  & 0      &0          &-17.4138  &0          &0\\
1  &$-$4.5071  & 0.0130 & 9.3846    & -23.7277 &0.4400  &1.7869\\
2  &   6.5080  & 3.2896 &-0.7031    & -10.0067 &1.2108 &2.5555\\
3  &  14.5596  & 3.3968 &2.1083     &-8.5767   &1.9001  &3.3284\\
4  &$-$14.2523 & 3.5523 &1.6296     &-15.5114  &2.5745  &4.6400\\
5  &$-$ 6.1681 & 3.6477 &4.4777     &-19.2964  &2.8057  &7.0698\\
6  &-- &-- &--                     &-8.4275    &2.9355  &9.8117\\
\end{tabular}
\end{ruledtabular}
\end{table}
%%%%%%%%%%%%%%%%%%%
%%%%%%%%%%%%%%%%%%%
%%%%%%%%%%%%%%%%%%%
%%%%%%%%%%%%%%%%%%%
\begin{figure*}[b!]
\includegraphics[scale=.62]{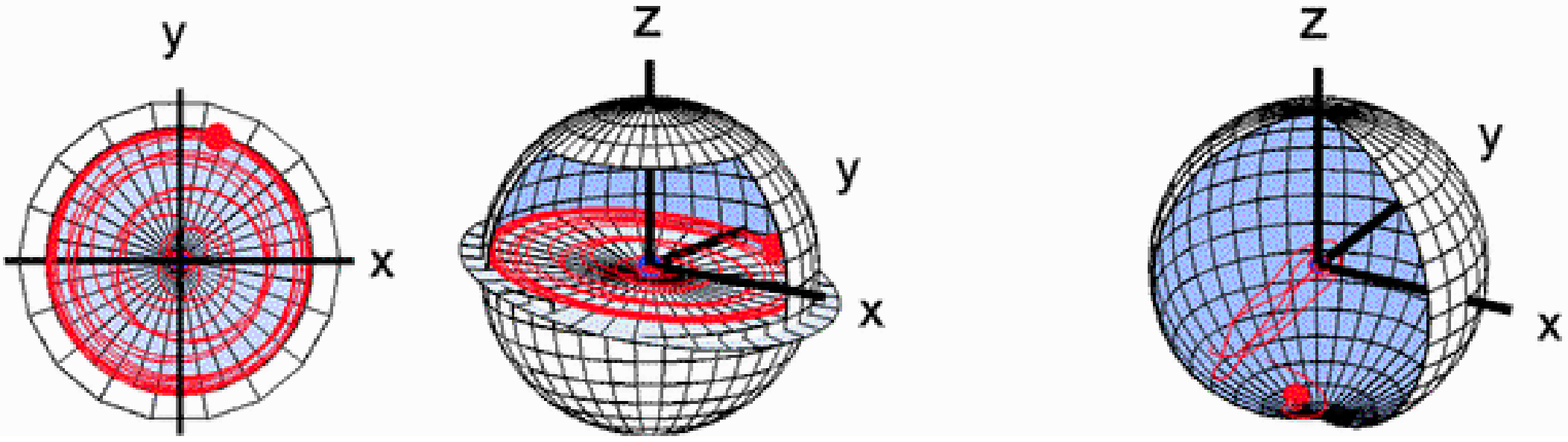}
\caption{Evolution of the Bell state 
$|\Phi_+\rangle = \tfrac{1}{\sqrt{2}}\left(|00\rangle+|11\rangle\right)$ 
as Fig.~2, but using the control of the experiment \cite{Nak03}.
Parameters of that pulse on the second qubit are: 
$\delta n_{g2}=0.25$, 255 ps total length, with 40 ps rise time and       
40 ps fall time, digitisation: 1000 points.
Note the different final states as compared with Figs.~1 and 2 indicative of
a resulting gate whose matrix elements coincide with the proper {\sc cnot} in absolute value, 
but not in phase. (Actually, the phase deviations are not uniform throughout the elements).
}
\end{figure*}
%%%%%%%%%%%%%%%%%%%
%%%%%%%%%%%%%%%%%%%
The representation in the Weyl chamber, Fig.\ \ref{fig:weyl} provide a complementary visualization. 
It represents the generic, irreducible two qubit part alone, i.e. each point in the chamber is invariant under single-qubit rotations. Details of this representation can be found in
Ref.\ \cite{KBG}. 
\begin{figure*}[h!]
\includegraphics{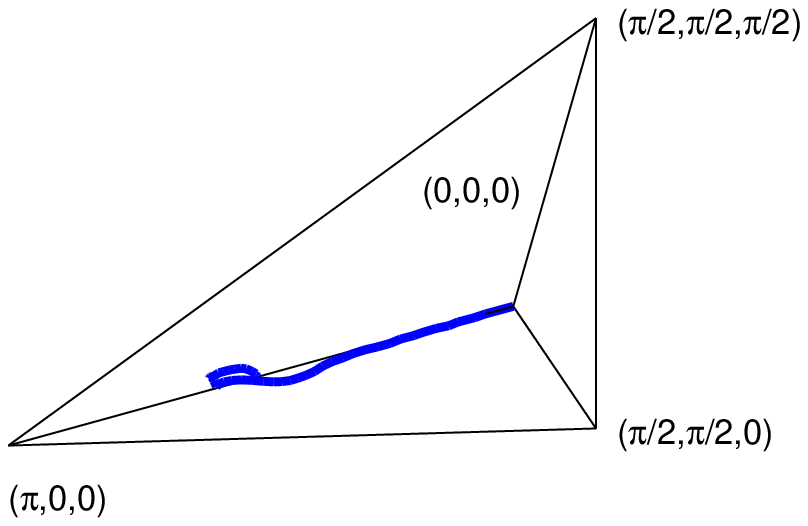}
\caption{Trajectory for the optimised {\sc cnot} in the Weyl chamber (compare ref.~\cite{KBG}). 
Starting at the origin, the trajectory is a non-geodesic smooth curve 
(see explanation in the main text) which 
ends at the point $(\tfrac{\pi}{2},0,0)$ as expected for {\sc cnot}. \label{fig:weyl}}
\end{figure*}
%%%%%%%%%%%%%%%%%%%
\section{Pulse shaping hardware}

\begin{figure*}[h!]
\begin{minipage}{0.49\columnwidth}
\includegraphics[width=0.9\columnwidth]{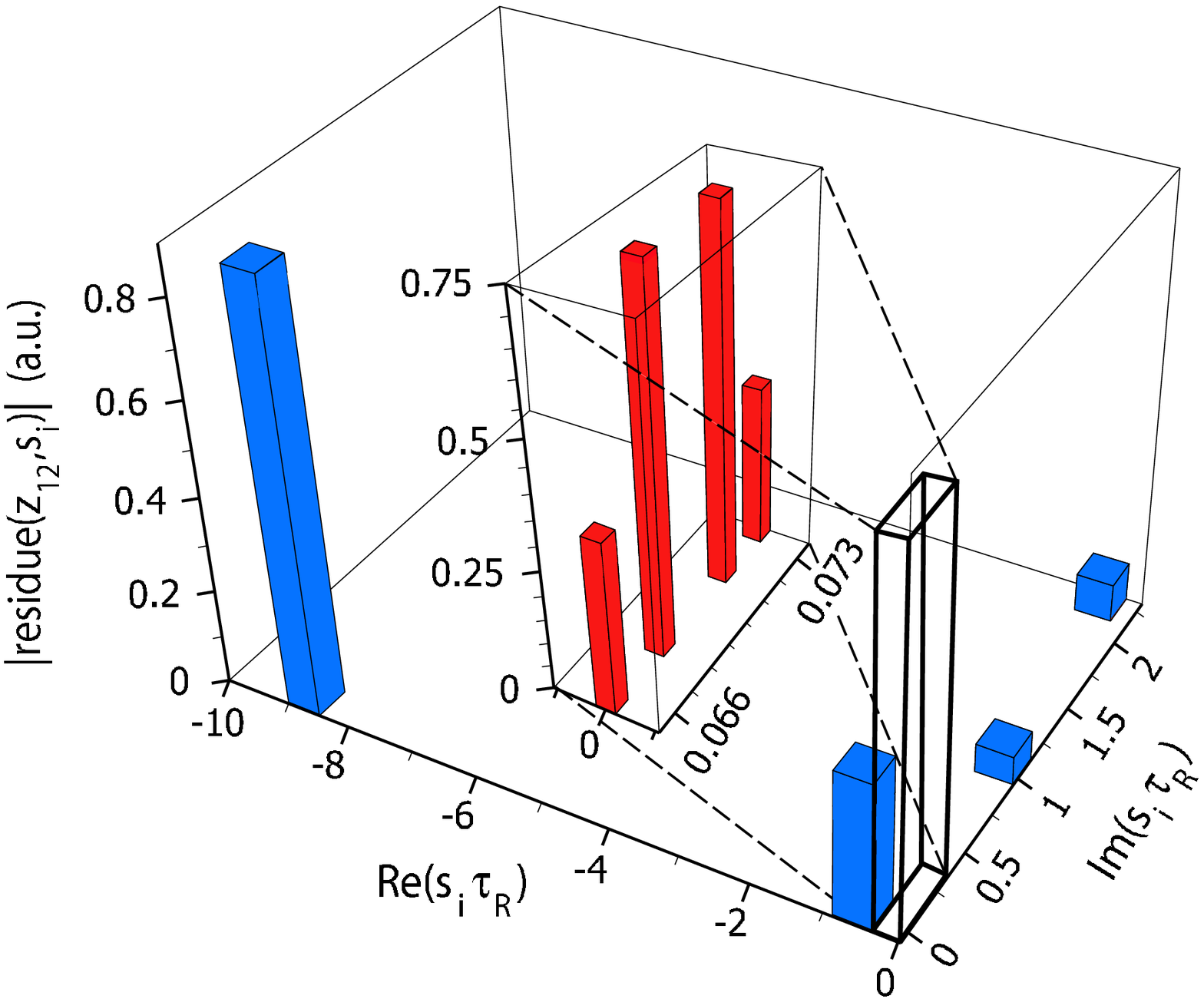}
\end{minipage}
\begin{minipage}{0.49\columnwidth}
\includegraphics[width=0.8\columnwidth]{arbeitsqubit_residua.eps}
\end{minipage}
\caption{Pole structure of the transfer functions necessary for shaping the pulses on both
the control (left) and target (right) qubit. Parameters correspond to Fig. 3 of the main paper. \label{fig:filter}}
\end{figure*}

This section details the pulse shaping scheme outlined in the main manuscript. The data in Fig.~3 
of the paper and in Fig.~\ref{fig:filter} of this supplement have been obtained as follows: We have fitted a rational function $Z_{12}(s)$ in Laplace space,
such that $V_{\rm out}(s)=Z_{12}(s)I_{\rm in} (s)$ where $I_{\rm in}$ is a $1 ps$ current pulse and
$n_{g,i}=C_{G,i}V_{\rm out,i}/2e$ for the two qubits, $i=1,2$. This function is represented best by its
residue decompostion $Z_{12}=\sum_i \frac{r_i}{s-s_i}$. With this decomposition, there are a number
of approaches to design a lumped circuit with this transfer function, such as the method of Gewertz \cite{Gewertz}. This methods systematically eliminates poles and introduces loops in the electrical 
circuit: An LCR-loop for each pair of complex conjugate poles, an RC-filter for each pole on the real 
axis. Thus, the degree of the polynomial in the denominator gives a clear view on the size of the necessary circuit.
In reality, the transfer function from the pulse shaping circuit, which can conveniently be placed 
at room temperature, to the sample is not smooth. By using a passive classical pickup element 
simulating the qubit, e.g. a capacitor in the charge qubit case \cite{Nak03}. This transfer 
function can be measured. In the linear case, it can be expressed as another four-pole 
impedance matrix $Z_{\rm sample}$. The
total transfer function of the series configuration of those four-poles will be 
$Z_{12}=Z_{12, {\rm sample}}Z_{12, {\rm filter }}/(Z_{22, {\rm filter}}+Z_{11, {\rm sample}})$. This outlines the statement in the text, that unless the transfer function to the sample is not filtering out the relevant
frequencies (i.e. becomes small for values of $s$ important to $V_{\rm out}$), it will be possible to design 
an appropriate filter taking into account the properties of the experimental setup.

\bibliography{draft}
%%%%%%%%%%%%%%%%%%%%%%%%
%%%%%%%%%%%%%%%%%%%%%%%%
%%%%%%%%%%%%%%%%%%%%%%%%